\documentclass{article}
\usepackage[dvips]{graphicx}
\usepackage{sprocl}

\def\be{\begin{equation}}
\def\ee{\end{equation}}
\def\bea{\begin{eqnarray}}
\def\eea{\end{eqnarray}}

\newcommand{\Corrftn}[2]{C_{\vec{#1}} (\vec{#2})}

\newcommand{\Source}[2]{S_{\vec{#1}} (\vec{#2})}
\newcommand{\Rftn}[2]{{\cal R}_{\vec{#1}} (\vec{#2})}

\newcommand{\wftn}[2]{\Phi^{(-)}_{\vec{#1}} (\vec{#2})}
\newcommand{\wfsquare}[2]{\left| \wftn{#1}{#2} \right|^2}

\newcommand{\etal}{{\em et al.}}                

\newcommand{\dn}[2]{{d}^{#1} #2 \:}

\begin{document}

\title{Imaging Three-Dimensional Relative Sources from Nuclear Reactions}

\author{DAVID A. BROWN}

\address{Institute for Nuclear Theory, University of Washington,
Box 351550\\ Seattle, WA 98195-1550, USA\\E-mail: dbrown@phys.washington.edu} 

\maketitle
\abstracts{ One can access the space-time development of a heavy-ion reaction 
directly by imaging the source function from two particle correlation 
functions.  In the case of like-charged pions, this imaging can be recast as a 
Fourier inversion problem.  We will demonstrate how this inversion can be 
performed on full three-dimensional (i.e. in long, side and out coordinates)
experimentally determined correlation functions.  We will discuss
the resulting three dimensional images of the relative sources.  Finally, 
we will discuss how to perform the full three dimensional inversion for 
particles whose final state interactions are more complicated than those of 
the pions.}
%
\section{Nuclear Interferometry}
%
Intensity interferometry has proven to be a valuable tool for nuclear physics
as it gives direct access to the space-time extent of heavy-ion reactions.
The correlation function, measured in interferometry,
typically is used either to extract the correlation radii or is compared to 
correlation functions generated from a semi-classical transport model.
Recently, it was noticed that one can make better use of the correlation 
function -- one can use it to {\em image} the relative source function of the
particles in question \cite{Brown:1997sn,Brown:1997ku}. 
This source function is the relative distribution of emission points of a
pair of particles in their center of mass (CM) frame.  The correlation radii
normally extracted from the correlation functions are the widths of this 
distribution.  While the extracted images represent an advance in the 
amount of information one can gather from correlation measurements, imaging 
was limited to angle averaged ($q_{inv}$) correlations.  We now demonstrate 
the imaging of full three dimensional (3D) correlation functions. 

In this talk, we will discuss how interferometry is really an inversion problem
and discuss two ways to solve the full 3D problem.  Both of these 
methods will be applied to a simulated Coulomb corrected pion correlation 
function.  The first method is a direct application of the Fast Fourier 
Transform (FFT) algorithm to the correlation data.  We will discuss the various
problems and limitations of this algorithm.  The chief limitation of this 
approach is that it only works for pion pairs.  The second method is a 3D 
implementation of the general imaging procedure used in 
\cite{Brown:1997sn,Brown:1997ku}.  Unlike the FFT method, this method works 
for all particle pairs.  Following the comparison of these two methods, we 
will outline future directions for the study of 3D correlation data.

%
\section{The Problem}

Our task is to extract as much information about the source function from the 
two particle correlation function as we can.  The experimently measured 
correlation function is defined as the following ratio:
\be
\Rftn{P}{q} =  \Corrftn{P}{q} - 1= \frac{dN_{2}/d^2\vec{p_1}\vec{p_2}}
{dN_{1}/d\vec{p_1} dN_{1}/d\vec{p_2}} - 1. 
\ee
As stated above, the source function is the relative distribution of emission 
points in the pair CM and in the Pratt-Koonin formalism it 
is \cite{Koonin:1977fh,Pratt:1990}
\be
\Source{P}{r}\equiv\int \dn{}{t_1}\dn{}{t_2}\int \dn{3}{R}
\;\sigma\left(\vec{R}+\vec{r}/2,t_1,\vec{P}=0\right)
\sigma\left(\vec{R}-\vec{r}/2,t_2,\vec{P}=0\right).
\ee
Here, $\sigma(\vec{r},t,\vec{p})\:d^3r\:dt$ is the probability for 
emitting one of the particles at time $t$ at position $\vec{r}$ with 
momentum $\vec{p}$.  One should note that all time dependence is integrated 
out of the source function.  For the purpose of this talk, the details of 
$\sigma$ are not important.

We can extract the source function directly from the data because
the source function and the correlation function are related through the
Pratt-Koonin Equation \cite{Koonin:1977fh,Pratt:1990}:
\be
\Rftn{P}{q} = \int \dn{3}{r} \left(\wfsquare{q}{r} - 1 \right) \Source{P}{r}
\equiv \int \dn{3}{r} \;K(\vec{q},\vec{r})\Source{P}{r}.
\label{eqn:pk}
\ee
Here $\wftn{q}{r}$ is the pair (anti-)symmetrized wave function in the pair 
CM frame.  One should note that this is a simple integral 
equation with a kernel $K(\vec{q},\vec{r})$.  In the next few sections we will 
demonstrate the different ways to invert this integral equation.

%
\section{Extracting the Source with the FFT}
%
In the case of pions, the task of imaging is simple.  Here, we can 
ignore final state interactions in the relative wavefunction, turning the 
problem into a Fourier transform problem.  When one does this, the kernel 
$K(\vec{q},\vec{r})$ becomes $K(\vec{q},\vec{r}) = 
\cos{(2\vec{q}\cdot\vec{r})} $ and we can immediately solve for the source:
\be
\Source{P}{r}=\frac{1}{\pi^3}\int\dn{3}{q}
   \cos{(2\vec{q}\cdot\vec{r})}\Rftn{P}{q}.
\label{eqn:FFTanswer}
\ee
This integral can then be performed with the Fast Fourier Transform algorithm.

The implementation of the FFT algorithm is simple.  First, we discretize 
eq.~(\ref{eqn:FFTanswer}), converting this equation to a finite Fourier 
transform.  Second, we perform the resulting transform with a canned FFT 
routine \cite{Press:1992}.  To compute the errors, we Monte Carlo sample the 
errors on the correlation function to generate a test correlation, invert the 
test correlation, then repeat.  After test 100 runs, we compute the standard 
deviation of the ensemble of test sources.

Now, while the FFT is fast, there are several problems we must deal with.
The first is dealing with statistical noise.  Typically this is remedied by 
using a filter such as the Weiner optimal filter \cite{Press:1992}.  
The second problem is that the data must 
have the number of points equal to a power of two.  In other words, we need to 
pad data to make the next power of two, artificially increasing the 
resolution.  The final problem is that the FFT approach only works for pion 
pairs.  

\begin{figure}
  \centering
  \includegraphics[angle=270,totalheight=2in,width=\textwidth]{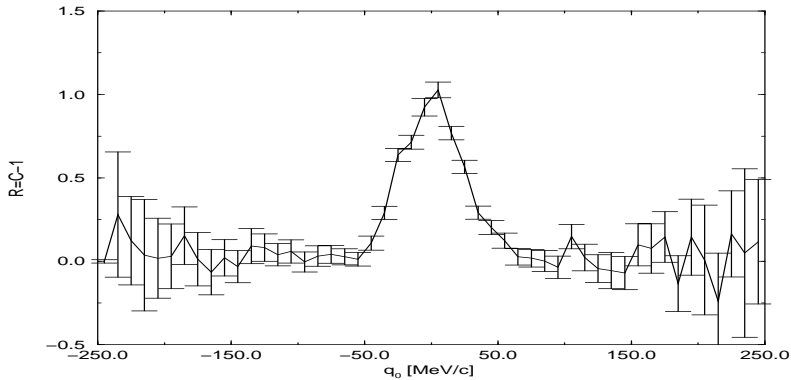}
  \caption[]{Sample plot of the model correlation function.  This plot is along
    the $q_L=q_S=0$ axis.}
  \label{fig:corr}
\end{figure}

\begin{figure}
  \centering
  \includegraphics[width=\textwidth]{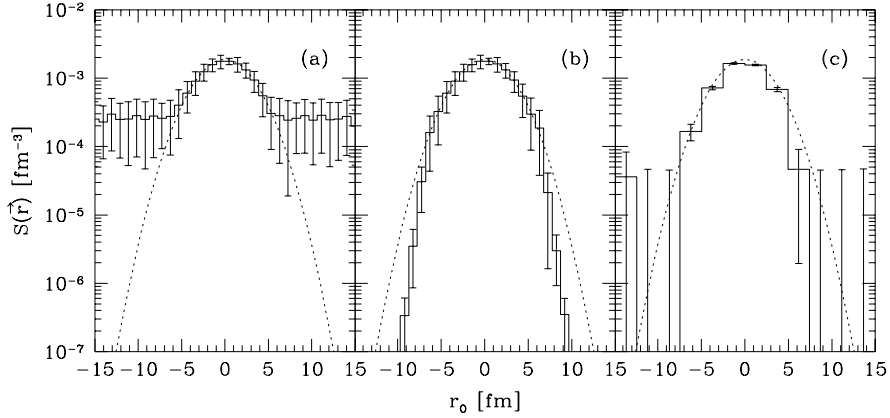}
  \caption[]{Sample plots of the source functions.  These plots are all along
    the $r_L=r_S=0$ axis.  In all three panels, the dashed curve is the input
    source.  In panel a., the source was imaged using the FFT algorithm alone.
    In panel b., the source was imaged using the FFT algorithm and the 
    filter described in the appendix.  In panel c., the source was imaged 
    using the general procedure.}
  \label{fig:sources}
\end{figure}

We can now test this inversion method.  To do this, we use a Gaussian 
test correlation with radius parameters $R_O=R_S=4$~fm and $R_L=6$~fm.  We 
take the error bars from a real data set and then add statistical noise to the
correlation to simulate an actual experimental (Coulomb corrected) 
correlation.  Fig.~\ref{fig:corr} is a sample of this full correlation 
function.  In fig.~\ref{fig:sources}a. we plot the results of the inversion
using the FFT alone.  As one can see, the tails overestimate the true source
by several orders of magnitude.  What is happening here is the FFT routine
is Fourier transforming the statistical noise in the data.   In 
fig.~\ref{fig:sources}b. we illustrate how the situation improves when one uses
the Weiner optimal filter \cite{Press:1992}.
While the tails are now much closer to 
the input source, they are now well below the input source.  In both cases, 
the inverted source {\em is not consistent with the input source}.  
A more sophisticated filtering method could likely fix this, but given the 
limitations of the FFT method, it is easier to abandon it entirely.  

%
\section{The General Approach}
%
The authors of \cite{Brown:1997sn,Brown:1997ku} give an approach to
inverting the Pratt-Koonin equation that is applicable for any particle 
pair and we will now outline it.  We seek the source that best represents
the experimental data.  Making a correlation out of a test source,
${\cal R}_i^{test}-1=\sum_j K_{ij}S^{test}_j$,
we would say that the test source reproduces the data well if 
it is the one that minimizes the $\chi^2$:
\be
\chi^2=\sum_j\left[\frac{({\cal R}^{exp}(q_j)-{\cal R}^{test}(q_j))}
  {\Delta {\cal R}^{exp}_j}\right]^2={\rm min}.
\ee
To find the minimum, we set $\delta \chi^2/\delta S_k = 0$ giving
a matrix equation for the source:
\be  
S_j=[(K^{\rm T}[\Delta^2{\cal R}^{exp}]^{-1}K)^{-1}K^{\rm T}
[\Delta^2{\cal R}^{exp}]^{-1}{\cal R}^{exp}]_j.
\ee
Similarly, by performing the error analysis one finds
the covariance matrix of the source:
$\Delta^2S_{ij}=[K^T [\Delta^2{\cal R}^{exp}]^{-1} K]^{-1}_{ij}$.

We apply this approach to the correlation described above.  In 
fig.~\ref{fig:sources}c., we plot the source imaged this way along with the
input source.  One can see that this source does as good a job at reproducing
the peak of the source as the FFT sources.  Also, this technique does
as good at reconstructing the tails as the FFT technique.  However, 
{\em unlike} the FFT, this technique returns errors that are much more 
realistic.  In other words, while the FFT claims it can image even below 
noise level, the general approach admits it can not image that well.
%
\section*{Final Comments}
%
We have learned several lessons by this comparison of the FFT and general
approaches.  While the FFT is fast it does not reliably report the 
uncertainty in the imaging.  On the other hand, the general method does 
reliably estimate the uncertainty.  Finally, while the FFT 
approach only works for pions, the general approach works for {\em any pair}.

There are many questions yet to answer using the 3D sources.
Do pions have a non-Gaussian tails due to resonances?  Proton one dimensional 
sources are not Gaussian, so what do they look like in 3D?
How does flow effect all 3D correlations? 
Given that the general method works for any pair, how about 3D
unlike pair correlations?  

As a final note, we invite readers to download 
and test our one dimensional inversion code.  It is available at:\\
{\centering http://www.phys.washington.edu/$\sim$dbrown/HBTprogs.html}
%
\section*{Acknowledgments}
%
I thank Pawe{\l} Danielewicz, Sergei Panitkin, John Cramer, 
Mike Lisa, and George Bertsch for discussions regarding this work.
\section*{References}
%


\begin{thebibliography}{99}


\bibitem{Brown:1997sn}
D.A.~Brown and P.~Danielewicz,
Phys. Rev. {\bf C57}, 2474 (1998).

\bibitem{Brown:1997ku}
D.A.~Brown and P.~Danielewicz,
Phys. Lett. {\bf B398}, 252 (1997).

\bibitem{Koonin:1977fh}
S.E.~Koonin,
Phys. Lett. {\bf 70B}, 43 (1977).

\bibitem{Pratt:1990}
S. Pratt \etal, Phys. Rev. {\bf C42} (1990).

\bibitem{Press:1992}
W. H. Press \etal,
{\em Numerical Recipes: The Art of Scientific Computation},  
Cambridge University Press, Cambridge, 1992.



\end{thebibliography}
\end{document}